\documentclass[fleqn,usenatbib]{mnras}

\usepackage{newtxtext,newtxmath}

\usepackage[T1]{fontenc}

\DeclareRobustCommand{\VAN}[3]{#2}
\let\VANthebibliography\thebibliography
\def\thebibliography{\DeclareRobustCommand{\VAN}[3]{##3}\VANthebibliography}

\usepackage{graphicx}	
\usepackage{amsmath}	

\usepackage{amssymb}	

\usepackage{mathtools}
\usepackage[dvipsnames]{xcolor}

\def\mywd{J1152$+$0248$-$V}
\def\mysystem{J1152$+$0248}

\title[The chemical structure of J1152+0248]{Uncovering the chemical structure of the pulsating low-mass white dwarf SDSS J115219.99+024814.4 }

\author[A. D. Romero et al. ]{
A. D. Romero$^{1}$\thanks{E-mail: alejandra.romero@ufrgs.br}, G. R. Lauffer$^{1}$, A. G. Istrate$^{2}$ and S. G. Parsons$^{3}$
\\
$^{1}$Physics Institute, Universidade Federal do Rio Grande do Sul, Av. Bento Goncalves 9500, Porto Alegre 91501-970, RS, Brazil \\
$^{2}$Department of Astrophysics, Radboud University Nijmegen, P.O. Box 9010, NL-6500 GL Nijmegen, the Netherlands\\
$^{3}$ Department of Physics and Astronomy, University of Sheffield, Sheffield, S3 7RH, UK
}

\date{Accepted XXX. Received YYY; in original form ZZZ}

\pubyear{2021}

\begin{document}
\label{firstpage}
\pagerange{\pageref{firstpage}--\pageref{lastpage}}
\maketitle

\begin{abstract}
Pulsating low-mass white dwarf stars are white dwarfs with stellar masses between 0.30~M$_{\odot}$ and 0.45~M$_{\odot}$ that show photometric variability due to gravity-mode pulsations. Within this mass range, they can harbour both a helium- and hybrid-core, depending if the progenitor experienced helium-core burning during the pre-white dwarf evolution.
SDSS J115219.99$+$024814.4 is an eclipsing binary system where both components are low-mass white dwarfs, with stellar masses of 0.362$\pm$0.014~M$_{\odot}$ and 0.325$\pm$0.013~M$_{\odot}$. In particular, the less massive component is a pulsating star, showing at least three pulsation periods of $\sim$1314 s, $\sim$1069 s and $\sim$582.9 s. This opens the way to use asteroseismology as a tool to uncover its inner chemical structure, in combination with the information obtained using the light-curve modelling of the eclipses.
To this end, using binary evolutionary models leading to helium- and hybrid-core white dwarfs, we compute adiabatic pulsations for $\ell=1$ and $\ell=2$ gravity modes with \texttt{Gyre}. We found that the pulsating component of the SDSS J115219.99$+$024814.4 system must have a hydrogen envelope thinner that the value obtained from binary evolution computations, independently of the inner composition. Finally, from our asteroseismological study, we find a best fit model characterised by T$_{\rm eff}=10\, 917$ K, M=0.338~M$_{\odot}$, M$_{\rm H}=10^{-6}$~M$_{\odot}$  with the inner composition of a  hybrid WD.

\end{abstract}




\section{Introduction}

White dwarfs (WDs) are the most common endpoint of stellar evolution. All stars with initial masses below 7$-$12~M$_{\odot}$ \citep[e.g.][]{1997MNRAS.289..973G, 2015ApJ...810...34W,2018MNRAS.480.1547L},  representing more than 95\% of the stars in the Milky Way, will end their lives as WDs.  The WD population can  be divided into hydrogen-rich atmosphere objects (DA), that correspond to more than 85\% of all WDs, and hydrogen-deficient objects (non-DA), which show no hydrogen in their atmospheres \citep[see][]{2008PASP..120.1043F, 2010A&ARv..18..471A}. 

For hydrogen-atmosphere WDs  the mass distribution peaks at $\sim$~0.6~M$_{\odot}$, which represents $\sim 84 \%$ of the total sample \citep{Kepler2007,2015MNRAS.446.4078K}, exhibiting also a high-mass and a low-mass components.
The low- and high-mass WDs are most likely the result of close binary evolution, where mass transfer and mergers commonly occur.

The low-mass tail in the DA mass distribution peaks at $\sim$~0.39~M$_{\odot}$ and extends to stellar masses lower than 0.45~M$_{\odot}$. White dwarfs with masses below $\sim$0.30~M$_{\odot}$ can only be formed through mass transfer in close binary systems, since single star evolution is not able to form such remnants in the Hubble time \citep{Kilic2007,2016A&A...595A..35I,2019MNRAS.488.2892P}. These objects are known as extremely low-mass white dwarfs (ELMs). Low-mass WDs are stars with stellar masses in the range of 0.30~$\leq$~M/M$_{\odot} \leq$~0.45. In addition to the binary formation channel, these objects could also form as a result of strong mass-loss episodes during giant stages for high metallicity progenitors. Noteworthy, low-mass WDs can harbour either a pure-helium core  \citep[e.g.][]{panei2007,althaus2013,istrate2014,2016A&A...595A..35I} or a hybrid core, composed of helium, carbon and oxygen  \citep[e.g.][]{iben1985, han2000, prada2009, zenati2019}. 

Probably the only way to probe the inner chemical composition in detail is through the pulsation period spectrum observed in variable stars. Each pulsation mode propagates in a specific region providing information on that particular zone, where its amplitude is maximum \citep{1990ApJS...72..335T}. Thus, we can perform an asteroseismic study, where we compare the observed periods with the theoretical period spectrum, computed using representative models, to uncover the inner stellar structure \citep[see e. g.][]{2012MNRAS.420.1462R,2019MNRAS.490.1803R}.

There are currently several families of pulsating WDs, that can be found in specific ranges of effective temperature and surface gravity. 
They show $g$-mode non-radial pulsations with periods range from minutes to a few hours and variation amplitudes of millimag.
The excitation mechanism acting on pulsating WDs is a combination of the $\kappa$-mechanism, driven by an opacity bump due to partial ionization of the main element in the outer layers \citep[][]{1981A&A...102..375D,1982ApJ...252L..65W}, and the $\gamma$-mechanism, related to the effect of a small value of the adiabatic exponent in the ionization zone \citep[][]{Brickhill1991,1999ApJ...511..904G}.

\begin{figure}
	\includegraphics[width=\columnwidth]{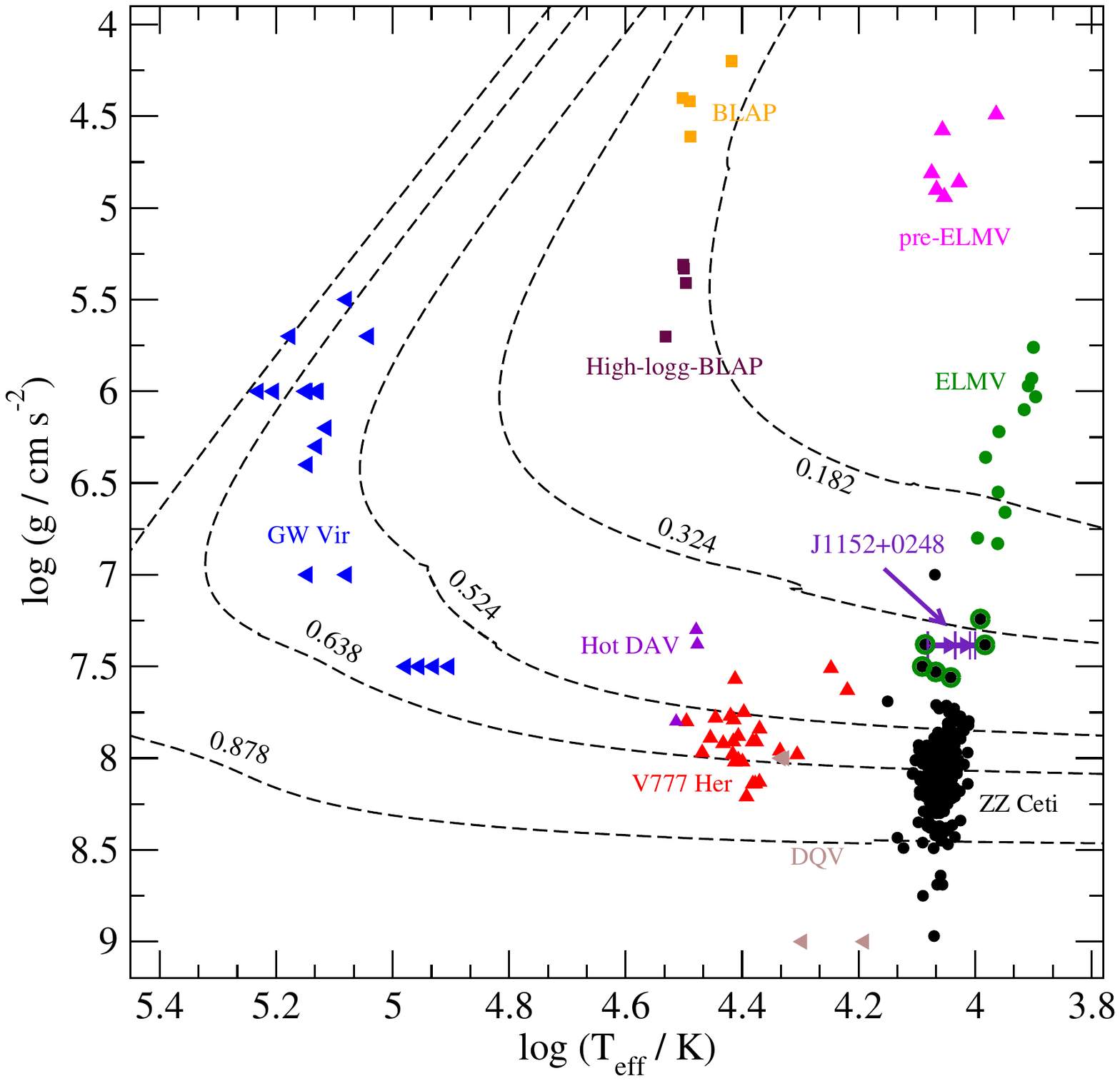}
     \caption{The classes of pulsating WDs. The data was extracted from \citet{2008PASP..120.1043F,2016IBVS.6184....1B,2017NatAs...1E.166P,2019A&ARv..27....7C,2019MNRAS.490.1803R,2019ApJ...878L..35K}. For reference,  we include theoretical WD sequences with C/O core and masses of 0.878, 0.638 and 0.524~M$_{\odot}$ \citep{2015MNRAS.450.3708R} and He-core with stellar mass of 0.324 and 0.182~M$_{\odot}$ \citep{2016A&A...595A..35I}. The different symbols indicate the element related to the excitation mechanism: hydrogen (circles), helium (triangle-up), carbon and/or oxygen (triangle-left) and iron peak elements (squares). The known variable low-mass WDs are indicated with a surrounding green circle. The position of J115219.99$+$024814.4 is depicted with a triangle-right, with the atmospheric parameters taken from \citet{Parsons2020} (see Table~\ref{tab:lm_observed_data} for details).  }
     \label{classes}
\end{figure}

The location of the different classes in the Kiel diagram is depicted in Figure \ref{classes}.
At high effective temperatures we find the GW Vir stars, with C/O atmospheres, followed by He-rich atmosphere V777 Her and the C-rich atmosphere DQV. Finally, the H-envelope pulsating WDs, known as ZZ Ceti stars, are located at lower effective temperatures.  
For lower $\log (g)$, we find the pulsating low-mass WDs, and the ELMs along with their progenitors,  the pre-ELMs. Even though the group of pulsating ELMs is considered a class on its own, their instability strip is an extension of the ZZ Ceti instability strip to lower surface gravities, as can be seen from Fig.~\ref{classes}.

There are 11 pulsating ELMs known to date \citep[green dots in Fig.~\ref{classes},][]{Hermes2013a, Hermes2013b,Kilic2015,Bell2015,Bell2017,2018MNRAS.478..867P}. The pulsating ELMs are characterised by periods in the range of $100 - 6\,300$~s, effective temperatures of $7\, 800 - 10\,000$~K and   a hydrogen dominated surface composition \citep{2019A&ARv..27....7C}. In addition, there are 10 objects in the literature with stellar masses within the range 0.30~$\leq$~M/M$_{\odot} \leq$~0.45 (black-green dots in Fig.~\ref{classes}) that show photometric variability with periods between 200 and 1300~s \citep{2016IBVS.6184....1B, Su+17,2017PhDT,Rowan+19}. For 4 of them, the uncertainties in the atmospheric parameters are quite large, leading to an uncertainty in stellar mass of 0.1--0.4~M$_{\odot}$ \citep{Su+17,Rowan+19}. The low number of pulsating low-mass WDs as compared to canonical mass ZZ Ceti stars could be due to some kind of fine tuning during the evolution of the progenitor, but it is most likely due to the lack of studies focused on the search for pulsations for objects in this stellar mass range. 

Recently, \citet{Parsons2020} reported the discovery of the first pulsating low-mass WDs in a compact eclipsing binary system (orbital period of  2.4 h), which happens to have another low-mass WD as a companion. The binary nature of the SDSS J115219.99$+$024814.4 system (hereafter \mysystem) was first reported by \citet{Hallakoun2016}, based on K2 data from the {\it Kepler} mission. 
\citet{Parsons2020} performed high-speed photometry observations with HiPERCAM on the 10.4 m Gran Telescopio Can\'arias in five different bands, with a total of 108 min of data, covering both primary and secondary eclipses. From the high time-resolution light curves they found pulsation-related variations from the cooler component with at least three significant periods. 
To determine the mass and radius of each component in \mysystem, they combine radial-velocity determinations from X-shooter spectroscopy with the information extracted from the primary and secondary eclipses in the light curves (see their Table 1).  The effective temperature was determined using two techniques, i.e. by fitting the spectral energy distribution  ($T_{\rm eff} {\rm (SED)}$) and by modelling the light curves including the effects of the eclipses ($T_{\rm eff} {\rm (Eclipse)}$). 
The stellar parameters obtained by \citet{Parsons2020} for the pulsating component in \mysystem~  (hereafter \mywd) are listed in Table \ref{tab:lm_observed_data}.

\begin{table}
	\centering
	\caption{Stellar parameters presented in \citet{Parsons2020} for the pulsating component of the \mysystem~ eclipsing binary.}
	\label{tab:lm_observed_data}
\begin{tabular}{cc}
\hline
   Parameter & Value \\
\hline
    M$_*$/M$_{\odot}$               & $0.325 \pm 0.013$    \\
    $R_*$/R$_{\odot}$               & $0.0191 \pm 0.0004$  \\
    $\log$ (g/[g cm$^{-2}$])           & $7.386 \pm 0.012 $  \\
    T$_{\mathrm{eff}}$/K   (SED) & $11\, 100 \pm^{950}_{770}$  \\
    T$_{\mathrm{eff}}$/K (Eclipse) & $10\, 400 \pm^{400}_{340}$  \\
\hline
\end{tabular}
\end{table}

Based on the determination of the radius, \citet{Parsons2020} proposed that \mywd~is either a hybrid-core or a helium-core low-mass WDs with an extremely thin surface hydrogen layer (M$_{\rm H}$/M$_{\odot} < 10^{-8}$). 
Different inner chemical structures will influence the characteristic period spectrum of a pulsating star. Therefore an asteroseismological study of this object can shed some light on both the chemical composition and the mass of the hydrogen envelope.

In this work, we explore the pulsational properties of both hybrid- and helium-core low-mass WD models representative for the case of \mywd. Furthermore, we consider sequences with hydrogen envelopes thinner than the value predicted by stable mass-transfer binary evolutionary models. For the evolutionary computations we use the stellar evolution code \texttt{MESA} \citep{MESA1,MESA2,MESA3,MESA4,MESA5}, while the  adiabatic pulsations  are computed using \texttt{GYRE} stellar oscillation code \citep{gyre1, gyre2}.  Using the theoretical period spectra, we perform an asteroseismological study of \mywd~to uncover its inner structure. 

This paper is organised as follows. In Section \ref{sec2}, we provide a description of the evolutionary computations and the input physics adopted in our calculations. In section \ref{sec:adiabatic_pulsations} we present the pulsation computations. Section \ref{results} is devoted to study the pulsational properties of our low-mass WD models, including a comparison between the hybrid and helium-core configurations. The results of the asteroseismological study of \mywd \ are presented in this section as well. Our final remarks are presented in Section \ref{sec:conclusions}.

\section{Evolutionary Sequences}
\label{sec2}

The evolutionary models presented in this work are computed using the open-source binary stellar evolution code \texttt{MESA}
\citep{MESA1, MESA2, MESA3, MESA4, MESA5}, version 12115  and are part of a grid of models covering the mass interval where helium- and hybrid-core WDs overlap, i.e. $\sim$0.32--0.45~M$_{\odot}$ (Istrate et al. 2021a, in preparation).  Since the aim  of this work is analyse the core-composition of \mywd \ using asteroseismology, we only consider sequences compatible with its observed effective temperature and radius.  We present evolutionary sequences for a WD mass of 0.325~M$_{\odot}$, corresponding to the value obtained by \citet{Parsons2020}, and 0.338~M$_{\odot}$ i.e. the maximum WD mass compatible within 1-$\sigma$. 

\subsection{Input physics} 
\label{sec21}
 We compute  binary evolutionary sequences using similar assumptions as in \cite{2016A&A...595A..35I}. All models include rotation, with the initial rotational velocity initialised such that the  donor is synchronised with the orbital period. We include magnetic  braking  for the  loss of angular momentum just for the donors leading to the formation of the helium-core WDs. For more massive donors ($>$2.0~ M$_{\odot}$), which are considered  for the progenitors of the hybrid-core WDs, we assume that the magnetic braking stops operating. This assumption follows from the    conventional thinking that braking via a  magnetized stellar wind is inoperative in stars with radiative envelopes \citep[][]{kawaler1988}.  For all the models, we assume  a mass transfer efficiency of 50 per cent, i.e. 50 per cent of the transferred mass is accreted by the companion, while the rest leaves the system with the specific angular momentum of the accretor.
 
Below, we briefly describe the main input physics considered in the evolutionary computations  and refer to Istrate et al. (2021a in preparation) for more details. 
We consider an initial metallicity of $Z=0.01$, with a helium abundance given by $Y=0.24+2.0\cdot Z$  and  the metal abundunces scaled according to \cite{grevesse1998}.
Convection is modelled using the standard mixing-length theory \citep{henyey1965} with a mixing-length parameter $\alpha$~=~2.0, adopting the Ledoux criterion. A step function overshooting extends the mixing region for 0.25 pressure scale heights beyond the convective boundary during core hydrogen burning. In order to smooth the boundaries we also include exponential overshooting  with $f=0.0005$. Semiconvection follows the work of \cite{langer1983}, with an efficiency parameter $\alpha_{sc}$ = 0.001. Thermohaline mixing is included with  an efficiency parameter of 1.0. Radiative opacities are taken from \cite{ferguson2005} for 2.7 $\leq$ log T  $\leq$ 3.8 and OPAL \citep{iglesias1993, iglesias1996} for 3.75 $\leq$ log T  $\leq$ 8.7, and conductive opacities are adopted from  \cite{cassisi2007}.

The nuclear network used is \textit{cno$\_$extras.net} which accounts for additional nuclear reactions for the CNO burning compared to the default \textit{basic.net}  network. We consider the effects of element diffusion \citep[e.g][]{iben1985,thoul1994} on all isotopes and during all the stages of evolution. 

We adopt a grey atmosphere using the Eddington approximation on the evolution prior to the cooling track, and the WD atmosphere tables from \cite{rohrmann2012} during the WD cooling stage. 

Rotational mixing and angular momentum transport are treated as diffusive processes as described in  \cite{heger2000}, with an efficiency parameter f$_{c}$ = 1/30  \citep{chaboyer1992} and a sensitivity to composition gradients parametrized by f$_{\mu}$ = 0.05. We also include transport of angular momentum due to electron viscosity \citep{itoh1987}.

\subsection{White dwarf formation history }
\label{subsec:artificial_WD}

\begin{figure}
	\includegraphics[width=1.0\columnwidth]{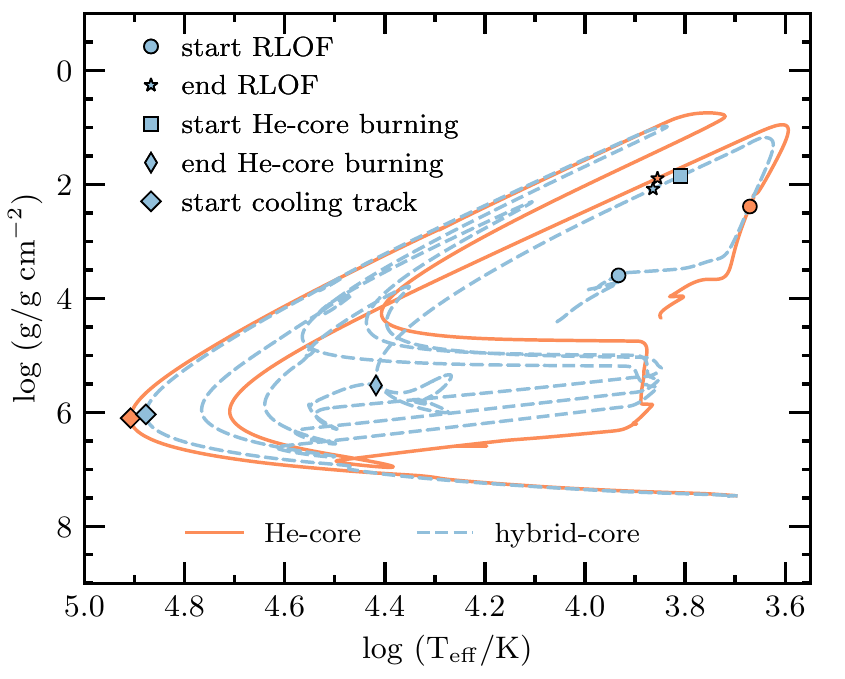}
    \caption{The Kiel diagram showing the formation and the cooling evolution of a 0.338~M$_{\odot}$ He- (solid orange line) and hybrid-core (dashed blue line) WD. The initial binary parameters are  M$_{\mathrm{donor}}$ = 1.3~M$_{\odot}$, M$_{\mathrm{accretor}}$ =1.2~M$_{\odot}$, P$_{\mathrm{initial}}$ = $\sim$16.982~days for the helium-core model, and  M$_{\mathrm{donor}}$ = 2.3~M$_{\odot}$, M$_{\mathrm{accretor}}$ =2.0~M$_{\odot}$, P$_{\mathrm{initial}}$ = $\sim$1.990~days for the hybrid-core model, at an initial metallicity of Z=0.01.  In both cases, the accretor is treated as a point mass.  The  evolutionary tracks are computed from ZAMS until the effective temperature of the WD reaches 5\;000~K. The symbols mark various stages of evolution. The beginning and the end of the mass-transfer phase  are represented by the circle and star symbol respectively, the beginning and the end of the core-helium burning phase (in the case of the hybrid-core sequence) are  depicted by the square and thin diamond symbol respectively,  and finally the diamond symbol represents the beginning of the cooling track, i.e. the point when the evolutionary track reaches its maximum effective temperature. }
    \label{fig:HR}
\end{figure}

Figure~\ref{fig:HR} shows the evolution of the donor star in the Kiel diagram  leading to a remnant mass of 0.338 M$_{\odot}$ with a helium- (solid orange line) and  hybrid-core (blue dashed line).  The evolution is computed from the zero-age main sequence (ZAMS) until the  remnant WD cools down to an effective temperature of 5\;000~K.  In both cases, the companion star is treated as a point mass. We also mark on the evolutionary sequences  several important points. The moment when the donor star overflow its  Roche-lobe (RLOF), marking the beginning of the mass-transfer phase, is depicted with a circle symbol,  the end of the mass transfer phase is marked with a star, the beginning  and the end  of the  core-helium burning phase are showed with the square  and  thin diamond symbol, respectively. Finally, the beginning of the cooling track  is represented by the diamond symbol. 

The helium-core WDs are formed by stripping mass when the donor star is on its red-giant branch. The binary system consists of a low-mass donor star with a initial mass of 1.3~M$_{\odot}$, a main-sequence companion of 1.2 M$_{\odot}$ and an initial orbital period  of $\sim$11.75 and $\sim$16.98~days, for the progenitor of the 0.325~M$_{\odot}$ and 0.338~M$_{\odot}$, respectively.   
After the end of the mass-transfer phase,  the remnant evolves through the so-called proto-WD phase, in which unstable hydrogen burning leads to the occurrence of at least one hydrogen flash. 

The initial binary configuration leading to the formation of the hybrid-core WDs is a 
2.3 M$_{\odot}$ intermediate-mass donor star with a companion of  2.0 ~M$_{\odot}$ and an orbital period of 1.43~ and $\sim$1.99~days, or the progenitor of the 0.325~M$_{\odot}$ and 0.338~M$_{\odot}$, respectively. The mass-transfer phase initiates during the Hertzsprung gap.  Unlike the case of the helium-core WD sequence, here the mass-transfer ceases due to the core-helium ignition. Shortly after the mass-transfer ended, the remnant starts the core-helium burning phase, which lasts around 670 Myrs. Once the CO-core is formed, the proto-WD undergoes four hydrogen shell flashes before finally settling on the cooling track. 


\subsection{White dwarf cooling track}
\label{sec:mass_radius}
The WD sequences discussed in the previous section are formed through a stable-mass transfer channel. While it is possible that the first mass-transfer phase which leads to the formation of \mywd~is stable, we cannot rule out completely a common-envelope evolution. Additionally, the   short orbital period of \mysystem~ (2.4~h) suggests that the second mass-transfer  phase leading to the formation of the most massive component is unstable and proceeds through a common-envelope evolution. This evolutionary phase could possibly also affect the lower-mass component. Either way, there is an uncertainty in the mass of the hydrogen envelope available at the beginning of the cooling track resulting from the evolutionary history prior to the  observed stage of the system.  

The mass of the hydrogen envelope is one of the  main factors that influences the cooling evolution of a WD. In order to take this uncertainty into account, we also computed WD cooling sequences  with hydrogen envelopes thinner than the ones obtained from the binary evolutionary models  described above. These sequences were computed using \textit{relaxation} methods available in MESA. The initial conditions are taken  at the point when the remnant reaches the beginning of the cooling track. Using the stellar profile at this point, we remove the desired amount of hydrogen, keeping the total mass unchanged (see Istrate et al. 2021b, for details).

\begin{figure}
	\includegraphics[width=1.0\columnwidth]{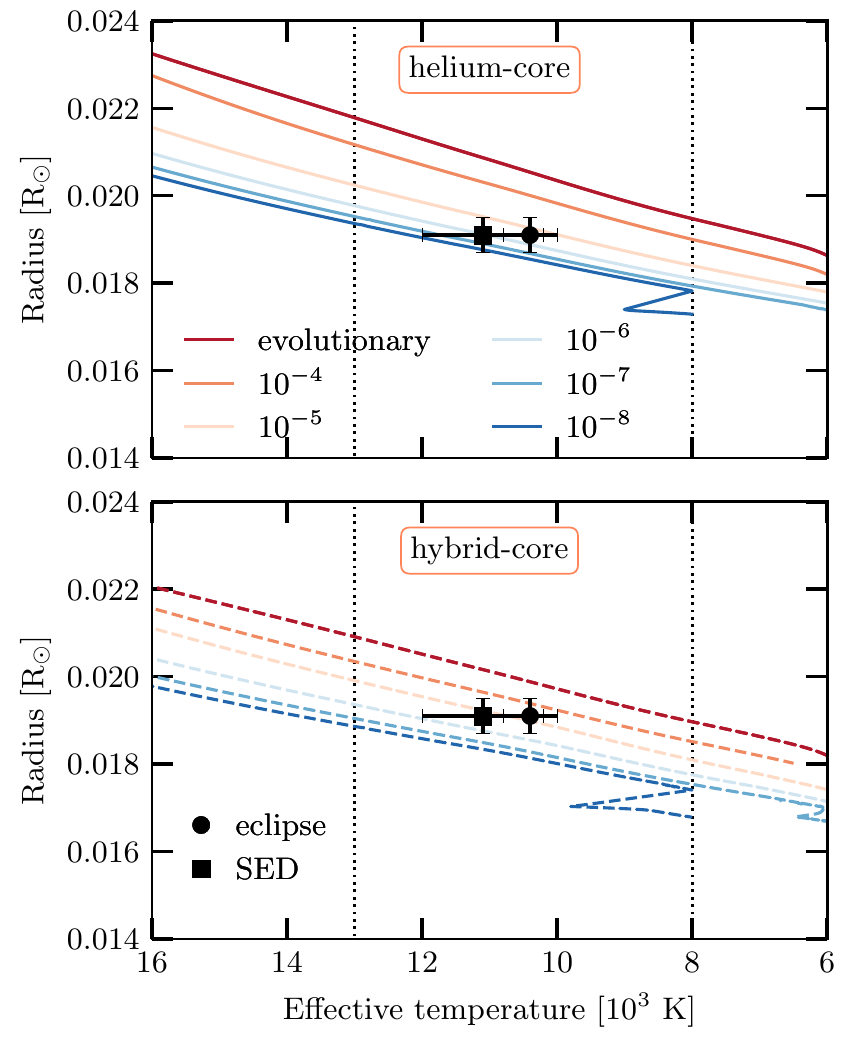}
    \caption{Stellar radius as a function of the effective temperature for both  hybrid- (dashed lines) and helium-core (solid lines) WD sequences  with stellar mass 0.338~M$_{\odot}$. The symbols correspond to the determinations of the effective temperature from observations (see table \ref{tab:lm_observed_data}). The dotted vertical lines indicate the blue and red edges of the observed instability strip for low-mass WDs (see Fig. \ref{classes}). }
    \label{fig:tracks}
\end{figure}

\begin{figure}
	\includegraphics[width=1.0\columnwidth]{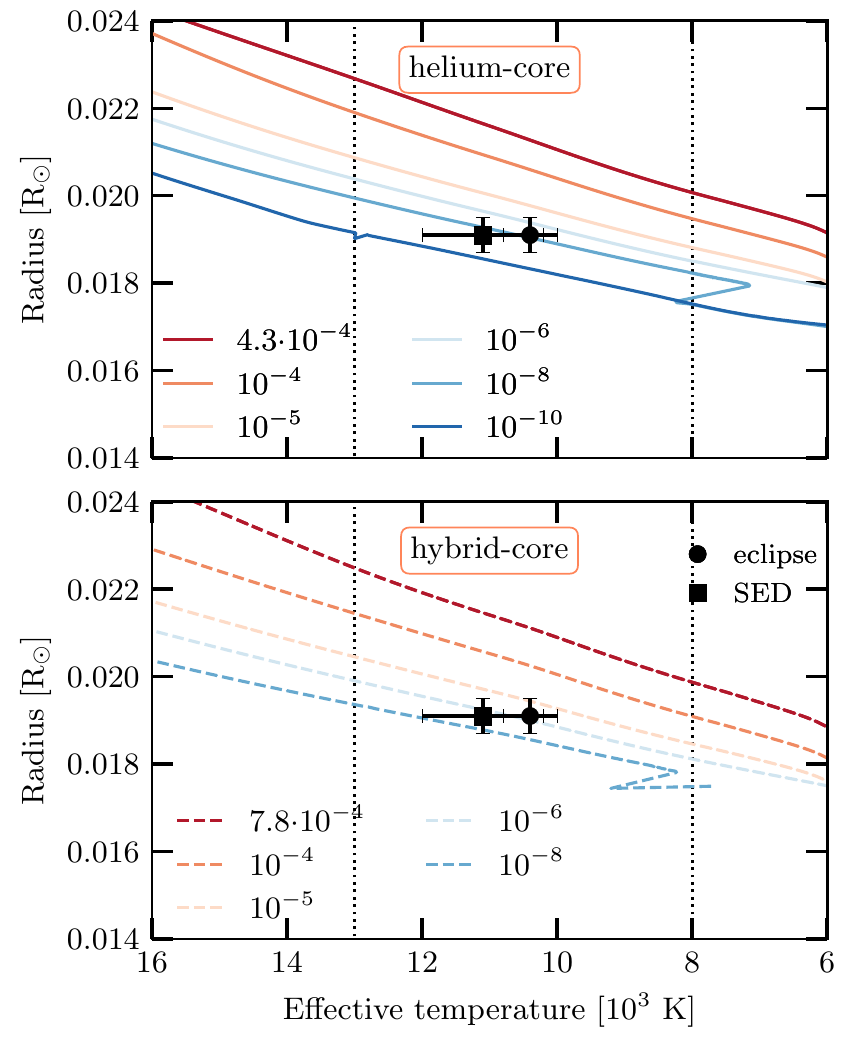}
    \caption{Same as Figure \ref{fig:tracks} but for the WD sequences with stellar mass 0.325~M$_{\odot}$.} 
    \label{fig:tracks1}
\end{figure}

Figures \ref{fig:tracks} and \ref{fig:tracks1} show the WD radius as a function of the effective temperature for sequences with  helium- (top panels) and hybrid-core (bottom panels) and stellar mass of 0.338~M$_{\odot}$ and 0.325~M$_{\odot}$, respectively.

For each core composition we consider different values of the hydrogen envelope mass, starting from the  value resulting from the binary evolution  down to 10$^{-8}$~M$_{\odot}$. In particular, for the stellar mass of  0.325~M$_{\odot}$ and helium-core, we also include a sequence with hydrogen-envelope mass of 10$^{-10}$~M$_{\odot}$.  Overplotted  are the measured values of \mywd, both from the eclipse  and the SED fitting. 

As expected, the radius for the hybrid-core sequences is smaller than for the helium-core sequences, for the same hydrogen envelope mass. 
Intriguingly, the observations are not compatible with a thick hydrogen envelope, i.e. that obtained from binary evolution computations.  For sequences characterised with a stellar mass of 0.338~M$_{\odot}$ (see Fig. \ref{fig:tracks}) the hydrogen-envelope mass consistent with the radius and effective temperature from \citet{Parsons2020} is between $10^{-4}$~M$_{\odot}$ and  $10^{-6}$~M$_{\odot}$ if we consider a hybrid-core. For the sequences with a helium-core, the hydrogen envelope mass is below $10^{-5}$~M$_{\odot}$.
For sequences with stellar mass of 0.325~M$_{\odot}$ (see Fig. \ref{fig:tracks1}), the hydrogen-envelope mass needs to be even smaller to fit the observations, between $10^{-5}$~M$_{\odot}$ and  $10^{-8}$~M$_{\odot}$ for sequences with a hybrid-core and $10^{-5}$~M$_{\odot}$ and $10^{-10}$~M$_{\odot}$ for sequences with a helium-core. Thus, \mywd~has a hydrogen envelope thinner than that obtained from evolutionary sequences, independently of the central chemical composition.

\section{Adiabatic Pulsations}
\label{sec:adiabatic_pulsations}

The pulsational properties in stars are dominated by the characteristic frequencies, i.e. the Brunt-V\"ais\"al\"a ($N^2$) and the Lamb ($L_{\ell}^2$) frequencies \citep{1941MNRAS.101..367C,1989nos..book.....U}. The Brunt-V\"ais\"al\"a frequency represents the oscillation frequency of a convective bubble around the stable equilibrium position. It is closely related with gravity modes since the restoring force is gravity. In the case of low-mass WDs,  the gravity modes correspond to the long period (low frequency) modes, with periods $\gtrsim 200$ s for $\ell=1$. The $N^2$ frequency is given by \citep{1991ApJ...367..601B}:

\begin{equation}
    N^2 = \frac{g^2\rho}{P}\frac{\chi_T}{\chi_{\rho}}\left(\nabla_{\rm ad}-\nabla+ B\right)
\end{equation}

\noindent where $g$, $\rho$ and $P$ are the gravitational acceleration, the density and the pressure, respectively. $\nabla_{\rm ad}$ and $\nabla$ are the adiabatic and the temperature gradient respectively, and $\chi_T$ and $\chi_{\rho}$ are defined as,

\begin{equation}
    \chi_T = \left(\frac{\partial \ln P}{\partial \ln T}\right)_{\rho}
\end{equation}

\begin{equation}
     \chi_{\rho} = \left(\frac{\partial \ln P}{\partial \ln \rho}\right)_{T}
\end{equation}

\noindent The term $B$ is called the Ledoux term and it is defined as,

\begin{equation}
    B=-\frac{1}{\chi_T}\left(\frac{\partial \ln P}{\partial ln X_i}\right)_{\rho,T}\frac{d\ln X_i}{d \ln P}
\end{equation}

\noindent where $X_i$ is the chemical abundance of the element $i$. The Ledoux term $B$ gives the explicit contribution of the change of chemical composition, thus it contributes in the region where chemical transitions are found \citep{1992ApJS...80..369B}. 

The Lamb frequency is given by:

\begin{equation}
    L_{\ell}^2 = \frac{(\ell +1)\ell}{r^2}c_S^2
\end{equation}

\noindent where $c_S$ is the adiabatic sound speed. The Lamb frequency is inversely proportional to the time that a sound wave takes to travel a distance $2\pi r/\ell$. It is closely related to pressure modes since the restoring force is the pressure gradient. Pressure modes correspond to short period (high frequency) modes with periods of $\lesssim 10$ s for low-mass WDs.

For a chemically homogeneous and radiative star, the period spectrum is characterised by a constant period separation, known as the asymptotic period spacing \citep{1990ApJS...72..335T},

\begin{equation}
    \Delta \Pi_a = \frac{\Pi_0}{[\ell(\ell+1)]^{1/2}}
    \label{Pasy}
\end{equation}

\noindent where

\begin{equation}
    \Pi_0 = 2\pi^2 \left(\int_{r_1}^{r_2}\frac{N(r)}{r}dr \right)^{-1}
    \label{pi0}
\end{equation}

In the case of WDs, the inner chemical structure is far from homogeneous, showing composition gradients \citep{2010ApJ...717..897A}. In this case, the asymptotic period spacing corresponds to the period spacing at the limit of very large values of $k$ \citep{1990ApJS...72..335T}.  

\subsection{Numerical computations}

For our pulsational computations we employed \texttt{GYRE} \citep{gyre1,gyre2}, an open--source stellar oscillation code in its adiabatic form, version 5.0. Although \texttt{GYRE} is integrated within \texttt{MESA}, it was used as an stand-alone package. We performed a scan using a linear grid over a period range of $80-2000$~s with \texttt{alpha\_osc}$=100$ and \texttt{alpha\_exp} and \texttt{alpha\_ctr} set to 50. The boundary conditions and variables were set to \text{UNNO} and \texttt{DZIEM}, respectively\footnote{The inlist and all its configuration settings can be found in  here [link to be updated]}. 

We compute pulsations for helium- and hybrid-core models in the effective temperature range of $13\, 000$ and $8\, 000$ K, which covers the empirical instability strip. We compute the period spectrum for $\ell = 1$ and $\ell = 2$ gravity modes with periods in a range of $80 \ {\rm s} \leq \Pi \leq 2\,000$~s.


\subsection{Internal composition and the characteristic frequencies}

\begin{figure}
	\includegraphics[width=0.95\columnwidth]{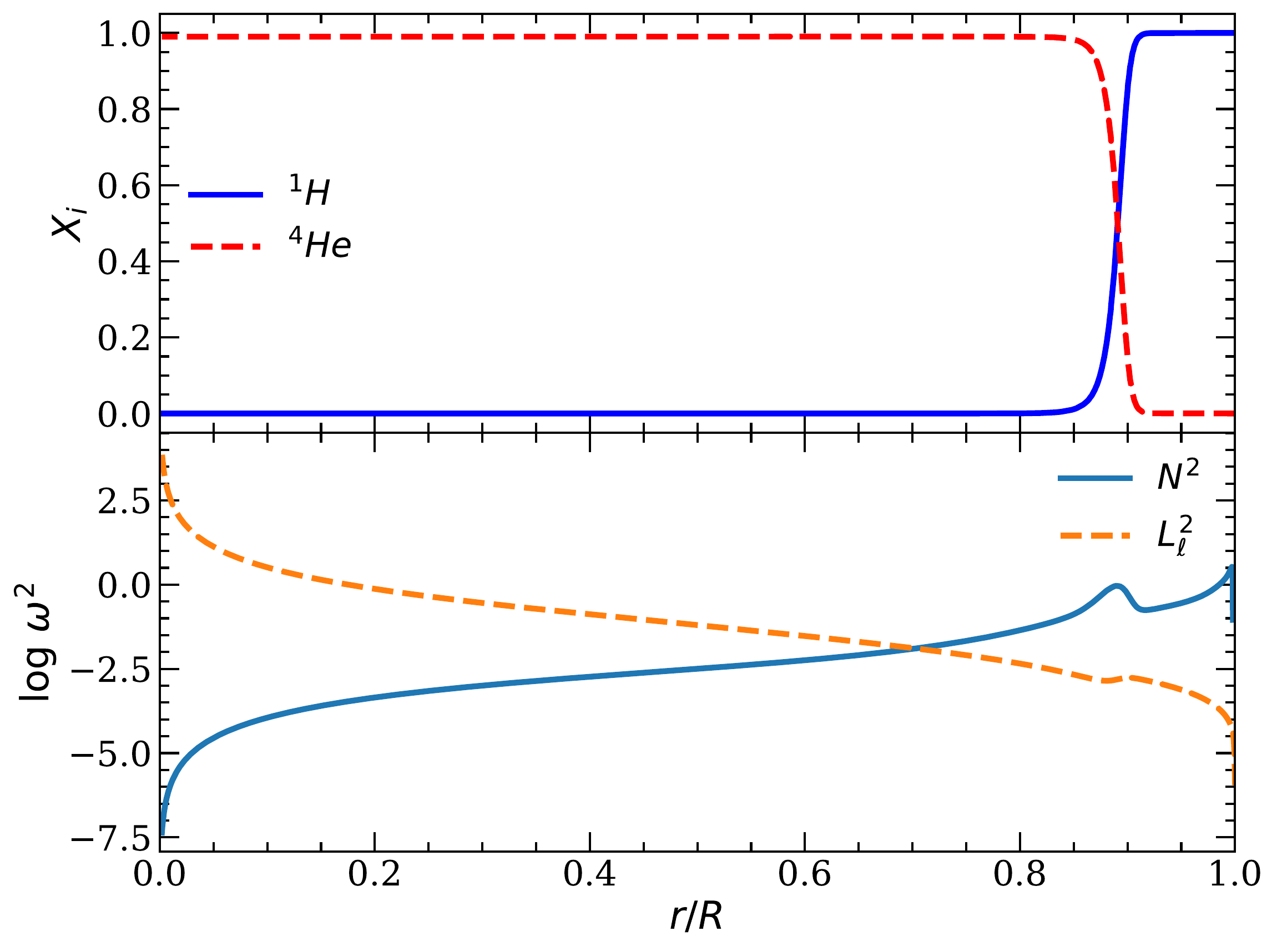}
    \caption{Internal profiles and characteristic frequencies for a \textit{helium-core} cooling sequence with M=0.338~M$_{\odot}$, T$_{\rm eff} = 10\, 000$~K and M$_{\rm H} = 10^{-4}$~M$_{\odot}$. \textit{Top} panel: the mass fraction of hydrogen and helium as a function of the relative radius. \textit{Bottom} panel: the logarithm of the Brunt-V\"ais\"al\"a (full blue line) and the Lamb (dashed orange line) characteristic frequencies as a function of the relative radius.  }
    \label{fig:profile-He}
\end{figure}

\begin{figure}
	\includegraphics[width=0.95\columnwidth]{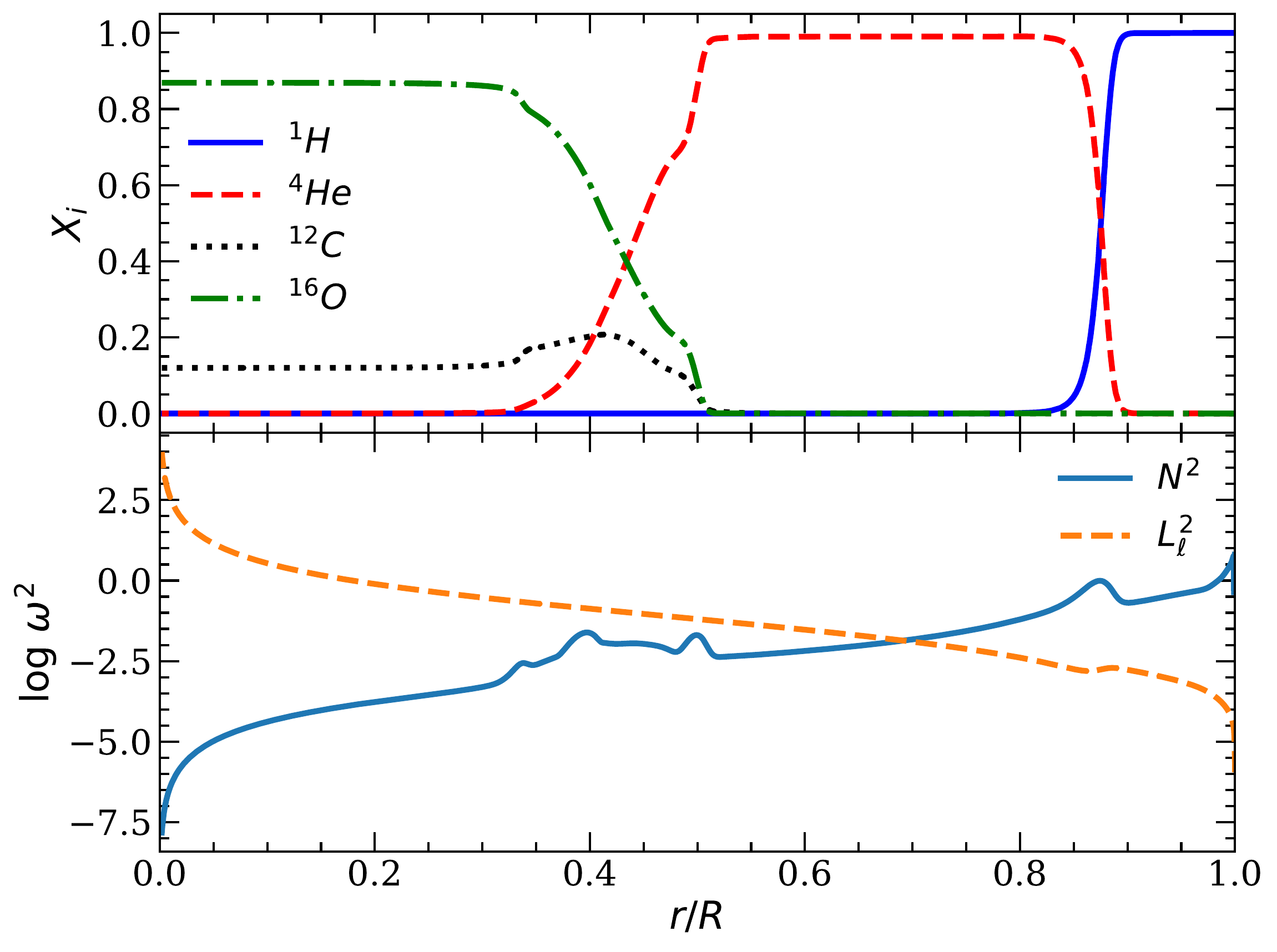}
    \caption{Internal chemical composition and characteristic frequencies for a \textit{hybrid-core} sequence with M=0.338~M$_{\odot}$, T$_{\rm eff} = 10\, 000$~K and M$_{\rm H} = 10^{-4}$~M$_{\odot}$. \textit{Top} panel: mass fraction of hydrogen, helium, carbon and oxygen as a function of the relative radius. \textit{Bottom} panel: the logarithm of the Brunt-V\"ais\"al\"a (full blue line) and Lamb (dashed orange line) characteristic frequencies as a function of the relative radius. }
    \label{fig:profile-Hy}
\end{figure}

To compare the pulsation properties of  helium- and hybrid-core WD models, we chose one template model for each central composition, with stellar mass of 0.338~M$_{\odot}$, effective temperature of $10\, 000$~K and hydrogen envelope mass of $10^{-4}$~M$_{\odot}$.
In the top panel of Figures \ref{fig:profile-He} and \ref{fig:profile-Hy} we depict the chemical profiles for a helium-core and a hybrid-core template models, respectively. Both template models show a pure hydrogen envelope, since gravitational settling had enough time to separate the elements in the outer layers. For the hybrid-core model we have a second chemical transition (He/C/O transition) around $r/R\sim 0.4$, where the helium abundance decreases towards the centre while the carbon and oxygen abundance increase. 

The presence of chemical transitions, where the abundances of nuclear species vary considerably in radius, modifies the conditions of the resonant cavity in which modes should propagate as standing waves. Specifically, the chemical transitions act as reflecting walls, trapping certain modes in a particular region of the star, where they show larger oscillation amplitudes. Trapped modes are those for which the wavelength of their radial eigenfunction  matches the spatial separation between two chemical transitions or between a transition and the stellar centre or surface \citep{BrassardIV, 1993ApJ...406..661B, 2002A&A...387..531C}.

The bottom panels of Figures \ref{fig:profile-He} and \ref{fig:profile-Hy} show the run of the logarithm of the Brunt-V\"ais\"al\"a and Lamb frequencies as a function of radius for the template models. The overall behaviour of $\log N^2$ is a general decrease with increasing stellar depth, eventually reaching small values in the deep core, as a consequence of degeneracy. As expected, the bumps in $\log N^2$ correspond to the chemical transition, where the chemical gradients have non-negligible values \citep{1990ApJS...72..335T}. 
For the helium-core template model, the Brunt-V\"ais\"al\"a frequency shows only one bump corresponding to the H/He transition at $r/R \sim 0.9$. 
For the hybrid-core template model, in addition to the bump at the base of the hydrogen envelope, the Brunt-V\"ais\"al\"a frequency shows a structure corresponding to the He/C/O transition around $r/R\sim 0.4$. This transition is wide and shows three peaks due to the structure of the chemical gradients. We expect that the different profiles for the Brunt-V\"ais\"al\"a frequency for helium-core and hybrid-core models will impact the pulsation properties, for example the period spectrum and the period spacing.

The Lamb frequency is only sensitive to the H/He transition at the bottom of the hydrogen envelope, as it is the case for ZZ Ceti stars \citep{2012MNRAS.420.1462R}. Thus, we do not believe the pressure modes could give information on the inner regions, if ever detected in low-mass WD stars.

\section{Results}
\label{results}

\subsection{Pulsational properties}

\begin{figure*}
	\includegraphics[width=0.9\textwidth]{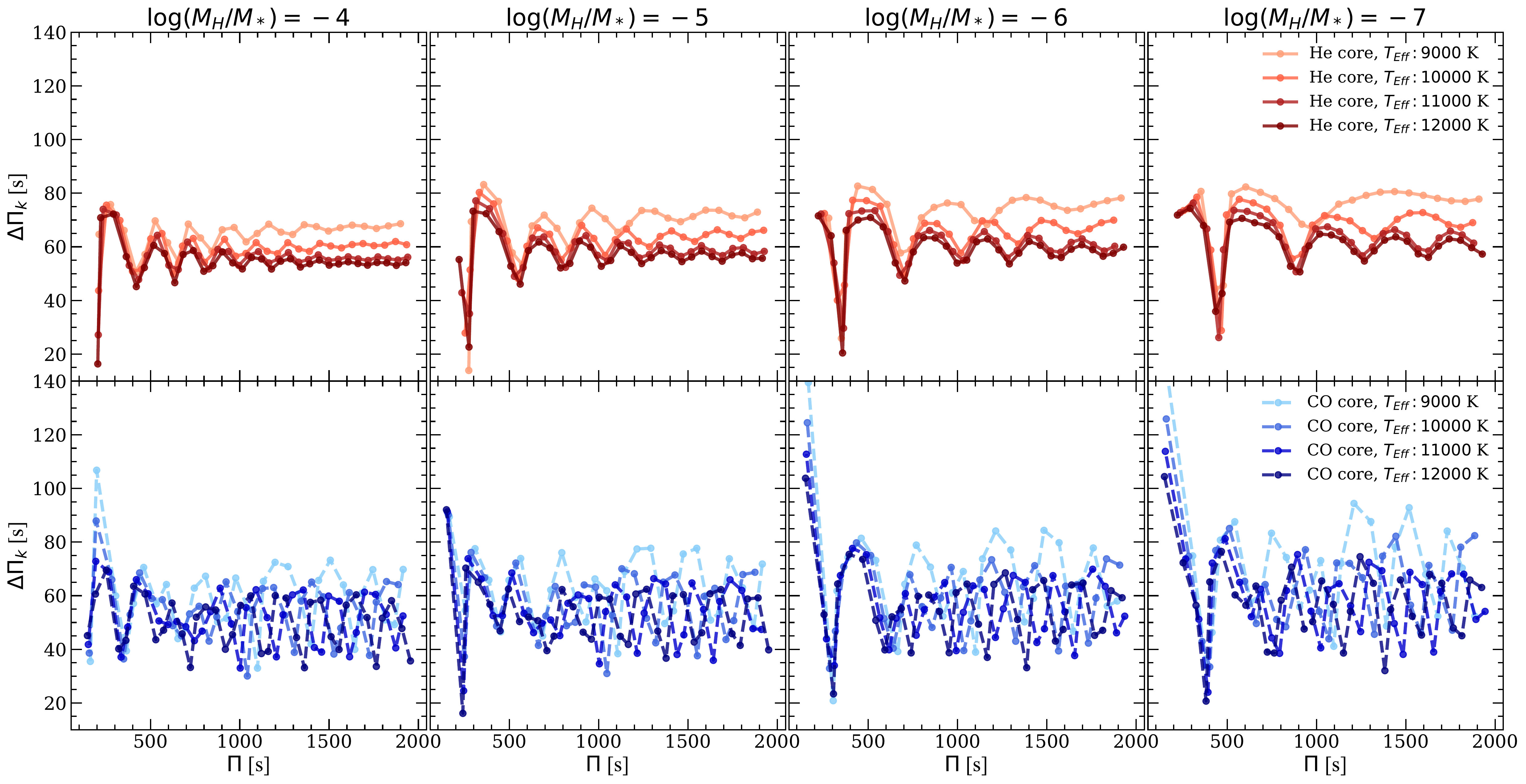}
    \caption{ The forward period spacing ($\Delta \Pi_k=\Pi_{k+1}-\Pi_k$) as a function of the period for modes with $\ell=1$ for helium-core (\textit{top} panels) and hybrid-core (\textit{bottom} panels) models, with stellar mass 0.338~M$_{\odot}$. Each column corresponds to a different hydrogen  envelope mass. In each plot we show the period spacing for four effective temperatures along the cooling sequence, $12\,000$, $11\,000$, $10\,000$ and 9000~K. }
    \label{fig:DP-l1}
\end{figure*}


As  previously mentioned, for a chemically homogeneous and radiative star, the forward period spacing ($\Delta \Pi_k=\Pi_{k+1}-\Pi_k$) would be constant and given by the asymptotic period spacing, defined in equation \ref{Pasy}. However, in the stratified inner structures, as the ones found in low-mass WDs, the forward period spacing deviates from a constant value, particularly for low-radial order modes.  In this section we will focus on models with stellar mass 0.338~M$_{\odot}$. Similar results are found for models with stellar mass of 0.325~M$_{\odot}$.

Figure \ref{fig:DP-l1} shows the forward period spacing as a function of the period for $\ell=1$ modes. Top panels correspond to models with helium-core while bottom panels show the results for hybrid-core models. Finally, each column corresponds to sequences with different values for the hydrogen-envelope mass and each curve shows the forward period spacing at a different effective temperature along the cooling curve, from $12\, 000$~K to 9000~K. 

For the helium-core models the distribution for the forward period spacing as a function of the period, shows a simple trapping cycle characteristic of one-transition models  \citep{BrassardIV, 2014A&A...569A.106C}, with defined local minima, as shown in the  top panels of Figure \ref{fig:DP-l1}. For the hybrid-core models (bottom panels in Fig.  \ref{fig:DP-l1}) the pattern in the forward period spacing is more complex largely due to the influence of the He/C/O transition in the Brunt-Vais\"al\"a frequency. Even though we expect the H/He transition to be dominant in the mode selection process, the presence of the second, broader, transition is not negligible. The differences in the pattern of the forward period spacing can be used to determine the inner composition of low-mass WDs, if enough consecutive periods are detected.

The trapping period, i.e. the period difference between two consecutive minima in $\Delta \Pi_k$, is longer for lower effective temperatures and thinner hydrogen envelopes. This effect is much  more evident for the helium-core models than for the hybrid-core models. 

In general, we expect the forward period spacing to increase with decreasing effective temperature and thinner envelopes \citep{BrassardIV}. This can be better explained in terms of the asymptotic period spacing, given by equation \ref{Pasy}. 
The higher values for $\Delta \Pi_a$, and  $\overline{\Delta \Pi}$, for lower effective temperatures results from the dependence of the Brunt-V\"ais\"al\"a frequency as $N\propto \sqrt{\chi_{T}}$, with $\chi_T \rightarrow 0$ for increasing degeneracy ($T\rightarrow 0$) \citep[see for instance][]{2012MNRAS.420.1462R}. In addition, the asymptotic period spacing, and the mean period spacing, are longer for thinner hydrogen envelopes \citep{1990ApJS...72..335T}.

If the stellar mass is fixed, the asymptotic period spacing for the helium-core model is longer than that for the hybrid-core model, for the same effective temperature and hydrogen envelope mass. This is related to the dependence of the  Brunt-V\"ais\"al\"a frequency on the surface gravity $N\propto g$, where $g\propto$~M/R$^2$. In this case, the radius is smaller for the hybrid-core models which  leads to a larger value of the surface gravity (see Figs. \ref{fig:tracks} and \ref{fig:tracks1}).

\begin{figure*}
\includegraphics[width=0.9\textwidth]{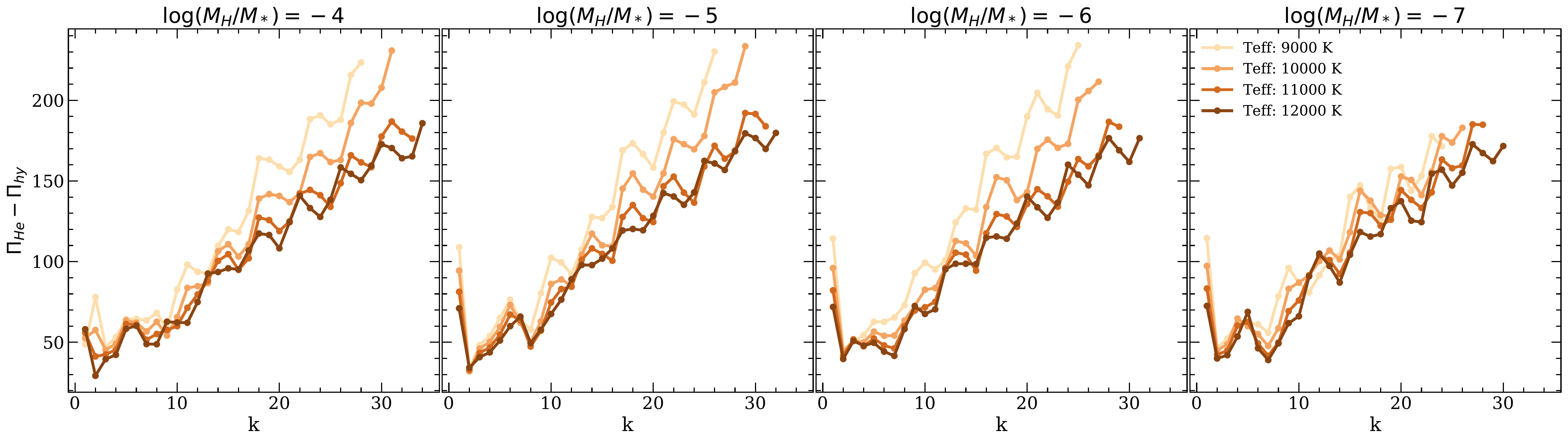}
\caption{Period difference ($\Pi_{\rm he}- \Pi_{\rm hy}$) as a function of the radial order for $\ell =1$ modes. The stellar mass is fixed to 0.338~M$_{\odot}$.} Each plot corresponds to a different hydrogen envelope mass. We consider four effective temperatures.
\label{fig:P-l1}
\end{figure*}


Finally, given the different chemical structure of helium- and hybrid-core WD models, we expect different values for the period spectra as well. 
Figure \ref{fig:P-l1} show the period difference between helium- and the hybrid-core models, $\Pi_{\rm he} - \Pi_{\rm hy}$ as a function of the radial order $k$, for harmonic degree $\ell=1$. Each column corresponds to a different hydrogen envelope mass, while the curves in each panel correspond to four different effective temperatures along the  0.338~M$_{\odot}$ WD cooling track. 
In all cases, the difference in the periods between the helium- and hybrid-core models are between $\sim 50$ s, for $k\lesssim 10$ and short periods, and $\sim 200$ s for $k\gtrsim 30$ and longer periods.  Note that the period difference is always positive, meaning that, for a given radial order, the period for the helium-core model is larger than the one corresponding to the hybrid-core model. This is expected, since the periods, and the period spacing, vary as the inverse of the Brunt-V\"ais\"al\"a frequency, and thus the surface gravity (Eq. \ref{pi0}). 
As the period also increases with  the cooling age, the period difference is larger for lower effective temperatures, particularly for modes with radial order $k> 20$. 
For future reference, we list the periods for $\ell=1$ and $\ell =2$, corresponding to the models considered in Figure \ref{fig:P-l1} in tables \ref{apendix1}, \ref{apendix2}, \ref{apendix3} and \ref{apendix4}.

\subsection{Asteroseismology of \mywd }
\label{star}

For \mywd~\citet{Parsons2020} detected three pulsation periods, with the period of $\sim 1314$ s being the one showing the highest amplitude. The list of observed periods and their corresponding amplitudes are listed in Table \ref{tab:perobs}.

\begin{table}
	\centering
	\caption{Observed pulsation periods and the corresponding amplitudes for \mywd~\citep{Parsons2020}.}
	\label{tab:perobs}
\begin{tabular}{cc}
\hline
Period [s] & amplitude [ppt] \\
\hline
$1314 \pm 5.9 $ & $33.0\pm 1.3$ \\
$1069 \pm 13  $ & $9.8\pm 1.3$  \\
$582.9 \pm 4.3$ & $8.9\pm 1.3$\\
\hline
\end{tabular}
\end{table}

\begin{figure*}
    \centering
    \includegraphics[width=\linewidth]{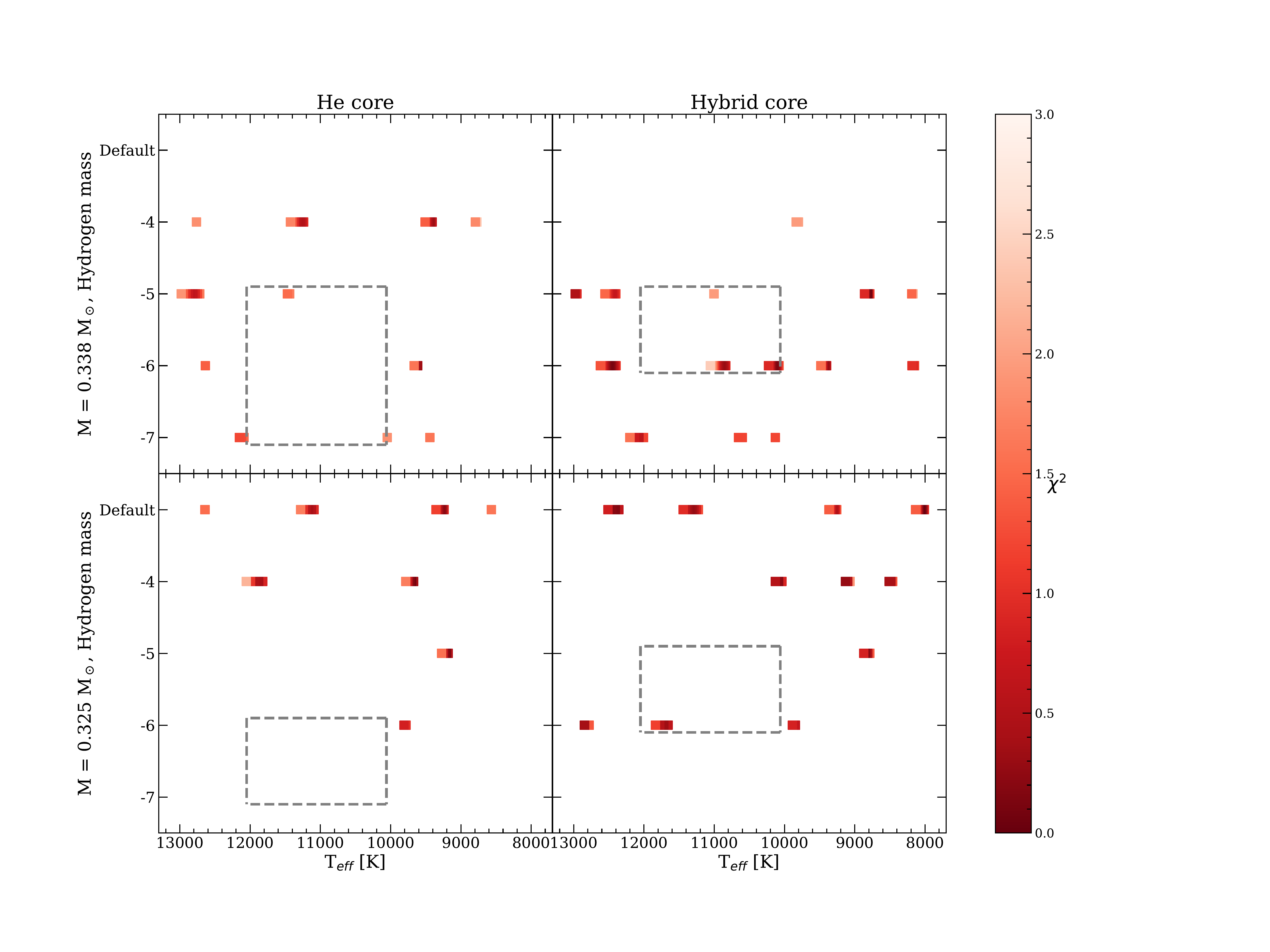}
    \caption{The value of $\chi^2$ (colour scale) as a function of the effective temperature and the mass of the hydrogen-envelope. Top panels correspond to models with stellar mass 0.338~M$_{\odot}$, while bottom panels show the results for 0.325~M$_{\odot}$ models. Both helium-core (left panels) and hybrid-core (right panels) structure are considered. The dashed-line rectangle indicates the region of the T$_{\rm eff}$- Hydrogen envelope plane restricted by the determinations of radius and effective temperature from \citet{Parsons2020}. }
    \label{bf}
\end{figure*}

In order to find the theoretical model that better matches the observed periods, we use a standard $\chi^2$ approach where we search for a minima of the quality function, defined as:

\begin{equation}
    \chi^2 = \sum_{i=1}^3 \left( \frac{ \Pi^{(i)}_{\mathrm{obs}} - \Pi^{(i)}_{\mathrm{th}}  }{\sigma^{(i)}_{\mathrm{obs}}} \right)^2
	\label{eq:chi2}
\end{equation}
where $\Pi^{(i)}_{\mathrm{obs}}$ and $\sigma^{(i)}_{\mathrm{obs}}$ are the observed periods and their associated uncertainties in $g_s$-band, and $\Pi^{(i)}_{\mathrm{th}}$ are the theoretical periods.
In our fit, we consider the mode with the highest detected amplitude, as an $\ell=1$ mode, since the amplitude is expected to decrease with increasing harmonic degree due to geometric cancellation \citep{1982ApJ...259..219R}.
For the remaining two observed periods we allowed them to be fitted by $\ell=1$ and $\ell=2$ modes. Finally, we restrict the possible seismological solutions to those models that fit the observed periods within the uncertainties reported by \citet{Parsons2020} (see Table \ref{tab:perobs}). The results of our seismological fit are depicted in Figure \ref{bf}. The value of $\chi^2$ (colour scale) is shown as a function of the effective temperature and the mass of the hydrogen envelope. Top (bottom) panels correspond to models with stellar mass 0.338~M$_{\odot}$ (0.325~M$_{\odot}$). Models with helium-core are depicted on the left panels, while the ones with hybrid-core are shown in the right panels. The dashed-line rectangle indicates the region where the models are compatible with the radius and effective temperature determinations from \citet{Parsons2020}. As can be seen from this figure, there are several families of solutions. For a stellar mass 0.325~M$_{\odot}$, the best fit models are mainly characterised by thick hydrogen envelopes, $\sim 10^{-4}$~M$_{\odot}$ and low effective temperatures,  $>9\, 000$ K, for both core compositions. Thin hydrogen-envelope solutions, with hydrogen mass below $\sim 10^{-5}$~M$_{\odot}$, are found for models with stellar mass 0.338~M$_{\odot}$ and hybrid-core, for all the effective temperature range. 

If we combine our seismological results with the restrictions from the radius and the effective temperature presented in \citet{Parsons2020} (dashed-line rectangle in Fig. \ref{bf}), the number of possible seismological solutions is largely reduced. In particular, the models characterised by a hybrid-core structure are the ones satisfying both criteria. 

\begin{table*}
	\centering
	\caption{Asteroseismological models for the low-mass WD \mywd. The effective temperature, hydrogen-envelope mass and central composition are listed in columns 2, 3 and 4, respectively. We list the theoretical periods that better fit the observed periods in column 4, along with the values of $\ell$ and k, in columns 6 and 7. The value of the quality function $\chi^2$ is listed in column 8. 	}
	\label{tab:bfm}
\begin{tabular}{cccccccccc}
\hline
\# & $T_{\rm eff}$ [K]& R$/$R$_{\odot}$ &M$_*$/M$_{\odot}$ & M$_{\rm H}$/M$_{\odot}$ & X$_{\rm C}$ &  $\Pi_{\rm th}$ [s] & $\ell$ & k & $\chi^2$ [s]\\
\hline
1 & $10\, 917$ & 0.0187 & 0.338 & $10^{-6}$ & C/O & 1315.4 & 1 & 21 & 0.296 \\
  &            &        &       &           &     & 1063.3 & 2 & 30 &   \\
  &            &        &       &           &     & 582.01 & 2 & 15 & \\
\hline
2 & $10\, 904$ & 0.0187 & 0.338 & $10^{-6}$ & C/O & 1316.1 & 1 & 21 & 0.305 \\
  &            &        &       &           &     & 1063.8 & 2 & 30 &   \\
  &            &        &       &           &     & 582.3 & 2 & 15 & \\
\hline
3 & $10\, 929$ & 0.0187 & 0.338 & $10^{-6}$ & C/O & 1314.9 & 1 & 21 & 0.318 \\
  &            &        &       &           &     & 1062.8 & 2 & 30 &   \\
  &            &        &       &           &     & 581.8  & 2 & 15 & \\
\hline
4 & $11\, 717$ & 0.0195 & 0.325 & $10^{-6}$ & C/O & 1316.1 & 1 & 21 & 0.345 \\
  &            &        &       &           &     & 1066.4 & 2 & 30 &   \\
  &            &        &       &           &     & 581.3 & 2 & 15 & \\
\hline
\end{tabular}
\end{table*}

The best fit models for \mywd~ are listed in Table \ref{tab:bfm}. These models fit the observed periods within the uncertainties and are in agreement with the determinations of radius and effective temperature presented by \citet{Parsons2020}. Note that all the models are characterised by a hydrogen-envelope mass of $10^{-6}$~M$_{\odot}$ and a hybrid-core. 
Thus, we can conclude that \mywd~ is a low-mass white dwarf star with a hybrid-core and a thin hydrogen-envelope.

\section{Conclusions}
\label{sec:conclusions}

In this work we explore the inner chemical structure of the pulsating low-mass white dwarf in the \mysystem~system using asteroseismology as a tool. This is the first pulsating low-mass WD in an eclipsing binary system with another low-mass WD which allows to constrain the mass and radius of  each component independently of evolutionary models.

We use low-mass WDs models with a stellar mass of 0.325 and 0.338~M$_{\odot}$ and a helium- and hybrid core respectively,  resulting from fully binary evolutionary computations.  In addition,  to account for the uncertainty in the hydrogen envelope mass due to the evolution through a common envelope channel, we use WD sequences  with the same stellar mass but with a hydrogen envelope thinner than the canonical value obtained from stable mass-transfer, i.e. from 10$^{-4}$ to 10$^{-10}$~M$_{\odot}$. For all these sequences, we calculate adiabatic pulsations for effective temperatures in the range $13\, 000$~K~$\leq T_{\rm eff} \leq$~8000~K. 

We perform a study on the pulsating low-mass WD in the binary system \mysystem~ to uncover its inner chemical composition. By comparing the determinations of the radius and effective temperature presented by \citet{Parsons2020} with theoretical models (Istrate et al. 2021b, in preparation), we find that the variable component of \mysystem~ must have a hydrogen envelope thinner than that predicted from stable-mass transfer  binary evolution computations. 

From the asteroseismological study we find a best fit model characterised by M$_*$=0.338~M$_{\odot}$, $T_{\rm eff}=10\, 917$ K, M$_{\rm H}$/M$_{\odot}=10^{-6}$ and a hybrid-core composition. In particular, all local minima of $\chi^2$ correspond to models with hybrid-core and a thin hydrogen-envelope. 

A systematic study of the pulsational properties and the observed period spectrum of low-mas WDs can give valuable information on the inner structure of these objects.  This will help in disentangling the two WD populations coexisting in the mass interval $\sim$0.32--0.45 M$_{\odot}$ which in turn leads to clues of the underlying progenitor population.

\section*{Data Availability}
 As an effort to promote open science, the results and codes to reproduce our results are available at \url{https://zenodo.org/communities/mesa/}.
 
\section*{Acknowledgements}

ADR and GRL acknowledges the support of the Coordena\c{c}\~ao de Aperfei\c{c}oamento de Pessoal de N\'ivel Superior - Brazil (CAPES) - Finance Code 001, and by Conselho Nacional de Desenvolvimento Cient\'ifico e Tecnol\'ogico - Brazil (CNPq). AGI acknowledges support
from the Netherlands Organisation for Scientific Research (NWO). SGP acknowledges the support of a Science and Technology Facilities Council (STFC) Ernest Rutherford Fellowship. Special thanks to Pablo Marchant, Josiah Schwab and Jocelyn Goldstein for their help with numerical debugging and to the developers of \texttt{MESA}  and \texttt{GYRE} for their continuous efforts to improve and extend these codes.    
This  research  has  made  use  of NASA’s Astrophysics Data System and open source software as the \texttt{Python 3} language \citep{Python2015}, the Ipython kernel \citep{ipython2007}, the Jupyter Project \citep{Jupyter2016} and the packages Matplotlib \citep{matplotlib2007}, Pandas \citep{pandas2010} and the Scipy ecosystem \citep{scipy2019}.




\bibliographystyle{mnras}
\bibliography{references}




\appendix

\section{Theoretical periods}

\begin{table}
    \centering
        \resizebox{\columnwidth}{!}{%
    \begin{tabular}{ cc|cc|cc|cc }
    \hline
  \multicolumn{2}{c}{$M_H/M_{\odot}$} & \multicolumn{2}{c}{$10^{-4}$} & \multicolumn{2}{c}{$10^{-5}$} & \multicolumn{2}{c}{$10^{-6}$} \\
     \hline 
$\ell$ & k &  He &  C/O &  He & C/O & He & C/O \\
        \hline
1 & 1  & 204.948 & 146.886 & 218.643 & 147.610 & 220.319 & 148.418 \\
1 & 2  & 221.260 & 192.002 & 273.896 & 239.660 & 291.871 & 252.223 \\
1 & 3  & 292.114 & 252.597 & 296.524 & 255.774 & 356.006 & 305.175 \\
1 & 4  & 364.331 & 322.131 & 369.816 & 326.066 & 376.420 & 328.547 \\
1 & 5  & 420.666 & 362.298 & 442.175 & 391.085 & 442.595 & 392.939 \\
1 & 6  & 465.821 & 405.630 & 507.842 & 447.835 & 512.597 & 468.370 \\
1 & 7  & 518.050 & 469.162 & 560.626 & 494.718 & 583.536 & 541.930 \\
1 & 8  & 578.586 & 529.785 & 606.713 & 557.217 & 650.957 & 592.739 \\
1 & 9  & 636.102 & 573.359 & 665.303 & 607.547 & 704.944 & 632.512 \\
1 & 10 & 682.723 & 620.481 & 727.082 & 659.543 & 752.211 & 684.640 \\
1 & 11 & 739.842 & 677.741 & 786.677 & 710.230 & 810.293 & 739.861 \\
1 & 12 & 798.462 & 723.525 & 838.796 & 749.740 & 873.722 & 778.522 \\
1 & 13 & 849.341 & 756.759 & 892.586 & 794.649 & 937.811 & 838.306 \\
1 & 14 & 902.294 & 808.744 & 954.758 & 856.940 & 997.247 & 898.565 \\
1 & 15 & 960.397 & 864.535 & 1014.63 & 912.796 & 1051.18 & 952.676 \\
1 & 16 & 1014.40 & 919.221 & 1067.36 & 959.175 & 1106.24 & 991.396 \\
1 & 17 & 1066.15 & 959.235 & 1122.26 & 1002.96 & 1168.11 & 1052.52 \\
1 & 18 & 1122.16 & 1004.73 & 1182.63 & 1062.39 & 1231.03 & 1116.87 \\
1 & 19 & 1177.61 & 1061.12 & 1240.57 & 1121.12 & 1289.88 & 1166.22 \\
1 & 20 & 1229.32 & 1121.06 & 1294.31 & 1165.97 & 1343.50 & 1203.20 \\
    \hline
2 & 1  & 120.014 & 100.316 & 141.344 & 100.940 & 142.736 & 101.493 \\
2 & 2  & 142.587 & 111.277 & 160.508 & 139.347 & 185.489 & 150.999 \\
2 & 3  & 185.975 & 151.547 & 187.659 & 152.494 & 209.625 & 176.847 \\
2 & 4  & 225.657 & 191.494 & 231.584 & 194.754 & 232.453 & 195.717 \\
2 & 5  & 254.034 & 211.840 & 272.358 & 229.803 & 273.019 & 231.131 \\
2 & 6  & 281.506 & 237.353 & 307.956 & 260.380 & 314.097 & 273.528 \\
2 & 7  & 314.200 & 273.648 & 334.955 & 288.159 & 354.935 & 316.423 \\
2 & 8  & 349.925 & 308.696 & 364.899 & 326.241 & 319.257 & 347.570 \\
2 & 9  & 379.496 & 336.314 & 400.421 & 358.342 & 418.988 & 372.001 \\
2 & 10 & 408.479 & 365.660 & 435.919 & 386.702 & 449.058 & 401.706 \\
2 & 11 & 443.206 & 396.338 & 468.693 & 416.106 & 484.623 & 439.694 \\
2 & 12 & 475.033 & 425.943 & 497.913 & 452.092 & 521.580 & 476.215 \\
2 & 13 & 504.023 & 461.149 & 531.857 & 483.738 & 557.625 & 498.294 \\
2 & 14 & 536.607 & 486.195 & 567.956 & 503.483 & 590.826 & 524.268 \\
2 & 15 & 569.643 & 506.188 & 600.466 & 533.089 & 621.325 & 560.439 \\
2 & 16 & 599.506 & 538.593 & 630.662 & 570.281 & 655.268 & 596.169 \\
2 & 17 & 630.599 & 572.313 & 664.302 & 597.936 & 691.754 & 617.974 \\
2 & 18 & 663.345 & 594.209 & 699.097 & 619.426 & 727.368 & 649.520 \\
2 & 19 & 694.175 & 617.784 & 731.092 & 652.467 & 759.583 & 684.916 \\
2 & 20 & 724.441 & 652.464 & 762.392 & 689.291 & 791.026 & 716.668 \\
2 & 21 & 756.787 & 684.847 & 795.637 & 714.925 & 825.950 & 738.317 \\
2 & 22 & 787.930 & 709.047 & 829.417 & 737.077 & 861.764 & 774.935 \\
2 & 23 & 818.226 & 731.327 & 861.601 & 772.178 & 895.634 & 813.010 \\
2 & 24 & 849.940 & 764.288 & 893.184 & 810.806 & 927.753 & 848.125 \\
2 & 25 & 881.227 & 800.000 & 926.496 & 843.600 & 960.895 & 870.022 \\
2 & 26 & 911.804 & 833.115 & 959.922 & 863.817 & 995.764 & 902.046 \\
2 & 27 & 943.122 & 853.700 & 991.721 & 894.125 & 1030.54 & 939.114 \\
2 & 28 & 974.435 & 879.080 & 1023.75 & 930.100 & 1063.74 & 966.625 \\
2 & 29 & 1005.23 & 913.682 & 1057.21 & 956.318 & 1096.41 & 989.506 \\
2 & 30 & 1036.10 & 941.506 & 1089.98 & 978.538 & 1130.42 & 1026.26 \\
\hline
\end{tabular}
}
   \caption{Period values for $\ell=1$ and $\ell=2$  modes corresponding to models with M$_*=$~0.338~M$_{\odot}$, effective temperature of $12\, 000$~K and hydrogen envelope mass of 10$^{-4}$, 10$^{-5}$, and 10$^{-6}$~M$_{\odot}$.}
    \label{apendix1}
    \end{table}
    
\begin{table}
    \centering
    \resizebox{\columnwidth}{!}{%
    \begin{tabular}{ cc|cc|cc|cc }
    \hline
  \multicolumn{2}{c}{$M_H/M_{\odot}$} & \multicolumn{2}{c}{$10^{-4}$} & \multicolumn{2}{c}{$10^{-5}$} & \multicolumn{2}{c}{$10^{-6}$} \\
     \hline 
$\ell$ & k &  He &  C/O &  He & C/O & He & C/O \\
        \hline    
1 & 1  & 207.777 & 151.919 & 233.957 & 152.645 & 235.524 & 153.322 \\
1 & 2  & 234.935 & 193.714 & 276.833 & 243.947 & 307.762 & 266.071 \\
1 & 3  & 308.929 & 266.516 & 311.893 & 268.497 & 361.733 & 309.919 \\
1 & 4  & 380.793 & 335.386 & 389.048 & 342.310 & 391.382 & 343.863 \\
1 & 5  & 433.845 & 372.487 & 463.274 & 408.527 & 463.749 & 411.409 \\
1 & 6  & 481.618 & 420.763 & 528.221 & 461.074 & 537.098 & 489.063 \\
1 & 7  & 538.336 & 486.755 & 577.219 & 513.621 & 610.573 & 564.377 \\
1 & 8  & 602.632 & 547.503 & 629.465 & 582.118 & 676.213 & 616.106 \\
1 & 9  & 655.766 & 598.093 & 692.730 & 635.549 & 725.608 & 655.982 \\
1 & 10 & 707.296 & 647.314 & 756.426 & 681.750 & 779.270 & 707.562 \\
1 & 11 & 769.074 & 697.705 & 814.497 & 731.479 & 843.416 & 768.330 \\
1 & 12 & 825.765 & 746.133 & 866.850 & 782.429 & 909.524 & 813.564 \\
1 & 13 & 877.491 & 789.346 & 928.721 & 827.532 & 973.853 & 868.318 \\
1 & 14 & 936.544 & 836.128 & 992.887 & 884.697 & 1032.09 & 927.724 \\
1 & 15 & 994.745 & 890.138 & 1049.75 & 944.978 & 1087.01 & 992.624 \\
1 & 16 & 1047.83 & 952.948 & 1104.48 & 1003.92 & 1149.66 & 1032.11 \\
1 & 17 & 1104.56 & 1002.48 & 1166.22 & 1038.52 & 1215.34 & 1085.88 \\
1 & 18 & 1162.76 & 1035.48 & 1227.62 & 1092.55 & 1277.54 & 1149.55 \\
1 & 19 & 1216.77 & 1091.03 & 1283.50 & 1156.59 & 1333.40 & 1211.86 \\
1 & 20 & 1272.22 & 1153.26 & 1340.63 & 1216.10 & 1392.04 & 1256.28 \\
\hline
2 &  1 & 120.711 & 103.726 & 149.787 & 104.375 & 151.814 & 104.829 \\
2 &  2 & 151.834 & 112.262 & 162.269 & 141.279 & 194.448 & 158.655 \\
2 &  3 & 195.703 & 159.298 & 197.606 & 160.112 & 211.759 & 179.466 \\
2 &  4 & 234.191 & 198.316 & 243.168 & 203.875 & 243.337 & 204.548 \\
2 &  5 & 261.843 & 218.039 & 284.536 & 239.630 & 285.963 & 241.754 \\
2 &  6 & 291.547 & 246.112 & 318.537 & 267.939 & 328.974 & 285.360 \\
2 &  7 & 327.190 & 283.652 & 345.423 & 299.045 & 370.474 & 328.794 \\
2 &  8 & 362.990 & 318.254 & 379.569 & 339.961 & 404.184 & 360.358 \\
2 &  9 & 390.988 & 350.257 & 416.755 & 375.420 & 431.958 & 388.493 \\
2 & 10 & 424.473 & 383.250 & 452.823 & 402.049 & 466.541 & 417.648 \\
2 & 11 & 459.804 & 409.554 & 483.925 & 429.733 & 504.598 & 454.562 \\
2 & 12 & 490.304 & 438.938 & 516.320 & 467.240 & 542.571 & 490.865 \\
2 & 13 & 521.968 & 475.075 & 553.706 & 498.686 & 578.574 & 513.862 \\
2 & 14 & 556.885 & 500.928 & 589.250 & 519.210 & 610.434 & 540.633 \\
2 & 15 & 588.750 & 521.371 & 620.508 & 550.428 & 644.063 & 580.230 \\
2 & 16 & 620.001 & 555.159 & 654.070 & 590.481 & 681.677 & 621.644 \\
2 & 17 & 653.702 & 590.195 & 690.425 & 625.170 & 719.126 & 649.981 \\
2 & 18 & 686.347 & 622.643 & 724.303 & 648.467 & 753.175 & 671.000 \\
2 & 19 & 717.267 & 644.476 & 756.359 & 674.238 & 785.390 & 707.347 \\
2 & 20 & 750.538 & 671.115 & 790.585 & 711.447 & 821.198 & 745.512 \\
2 & 21 & 783.051 & 704.597 & 825.706 & 743.906 & 858.469 & 768.119 \\
2 & 22 & 814.218 & 737.499 & 859.058 & 764.570 & 893.692 & 799.575 \\
2 & 23 & 846.959 & 757.240 & 891.796 & 797.047 & 926.893 & 838.324 \\
2 & 24 & 879.365 & 785.944 & 926.331 & 835.814 & 961.390 & 878.146 \\
2 & 25 & 911.008 & 822.649 & 960.885 & 872.099 & 997.639 & 907.292 \\
2 & 26 & 943.351 & 858.165 & 993.755 & 900.244 & 1033.59 & 933.155 \\
2 & 27 & 975.716 & 887.886 & 1027.15 & 924.922 & 1067.91 & 970.154 \\
2 & 28 & 1007.49 & 908.910 & 1061.95 & 960.303 & 1101.84 & 1008.85 \\
2 & 29 & 1072.09 & 940.675 & 1095.65 & 997.176 & 1137.52 & 1034.44 \\
2 & 30 & 1103.79 & 975.775 & 1128.77 & 1024.26 & 1173.68 & 1060.13 \\
\hline 
   \end{tabular}
   }
   \caption{Period values for $\ell=1$ and $\ell=2$  modes corresponding to models with M$_*=$~0.338~M$_{\odot}$, effective temperature of $11\, 000$~K and hydrogen envelope mass of 10$^{-4}$, 10$^{-5}$, and 10$^{-6}$~M$_{\odot}$.}     
    \label{apendix2}
\end{table}

\begin{table}
    \centering
        \resizebox{\columnwidth}{!}{%
    \begin{tabular}{ cc|cc|cc|cc }
    \hline
  \multicolumn{2}{c}{$M_H/M_{\odot}$} & \multicolumn{2}{c}{$10^{-4}$} & \multicolumn{2}{c}{$10^{-5}$} & \multicolumn{2}{c}{$10^{-6}$} \\
     \hline 
$\ell$ & k &  He &  C/O &  He & C/O & He & C/O \\
        \hline    
1 & 1  & 209.850 & 157.321 & 252.328 & 157.951 & 254.801 & 158.549 \\
1 & 2  & 253.518 & 195.924 & 280.175 & 248.189 & 326.917 & 283.059 \\
1 & 3  & 329.045 & 283.798 & 331.577 & 285.471 & 366.984 & 315.927 \\
1 & 4  & 398.872 & 349.747 & 411.843 & 361.596 & 412.719 & 362.727 \\
1 & 5  & 449.743 & 386.155 & 487.861 & 248.295 & 490.081 & 433.524 \\
1 & 6  & 501.311 & 439.089 & 550.052 & 476.945 & 567.284 & 513.238 \\
1 & 7  & 564.292 & 507.595 & 598.176 & 536.112 & 642.441 & 588.261 \\
1 & 8  & 629.238 & 566.539 & 658.379 & 608.287 & 703.193 & 639.718 \\
1 & 9  & 679.188 & 625.059 & 725.480 & 662.554 & 752.760 & 680.114 \\
1 & 10 & 739.471 & 673.875 & 790.184 & 704.021 & 815.755 & 733.170 \\
1 & 11 & 802.594 & 718.873 & 845.394 & 756.443 & 884.549 & 800.999 \\
1 & 12 & 856.823 & 772.083 & 904.980 & 819.879 & 953.196 & 856.930 \\
1 & 13 & 914.781 & 828.101 & 972.970 & 868.674 & 1017.90 & 905.083 \\
1 & 14 & 977.584 & 871.154 & 1036.10 & 918.761 & 1075.56 & 964.206 \\
1 & 15 & 1034.07 & 923.283 & 1092.89 & 982.771 & 1138.42 & 1034.76 \\
1 & 16 & 1091.60 & 988.449 & 1155.57 & 1045.98 & 1208.10 & 1074.30 \\
1 & 17 & 1153.16 & 1042.38 & 1222.19 & 1076.96 & 1277.21 & 1124.88 \\
1 & 18 & 1211.49 & 1072.43 & 1284.28 & 1129.61 & 1341.23 & 1190.77 \\
1 & 19 & 1268.95 & 1127.11 & 1344.28 & 1199.66 & 1402.32 & 1264.19 \\
1 & 20 & 1330.48 & 1189.77 & 1408.16 & 1267.85 & 1468.56 & 1325.45 \\
\hline        
2 &  1 & 121.586 & 107.287 & 157.518 & 107.919 & 163.156 & 108.312 \\
2 &  2 & 163.416 & 113.564 & 167.306 & 143.524 & 203.910 & 168.026 \\
2 &  3 & 206.981 & 168.965 & 209.618 & 169.896 & 216.181 & 183.062 \\
2 &  4 & 243.544 & 205.790 & 256.828 & 214.873 & 257.004 & 215.468 \\
2 &  5 & 271.659 & 226.439 & 298.421 & 250.869 & 302.091 & 254.650 \\
2 &  6 & 304.270 & 256.870 & 329.841 & 277.285 & 347.023 & 299.360 \\
2 &  7 & 343.101 & 295.672 & 359.639 & 312.151 & 388.134 & 342.354 \\
2 &  8 & 376.609 & 329.099 & 397.374 & 355.056 & 418.378 & 374.155 \\
2 &  9 & 406.703 & 365.975 & 436.177 & 392.490 & 450.250 & 406.368 \\
2 & 10 & 443.943 & 400.944 & 471.367 & 418.069 & 489.031 & 434.251 \\
2 & 11 & 478.181 & 424.514 & 502.886 & 445.030 & 529.062 & 470.434 \\
2 & 12 & 509.339 & 453.685 & 540.512 & 483.040 & 568.103 & 506.993 \\
2 & 13 & 545.056 & 489.279 & 579.302 & 514.825 & 603.410 & 532.999 \\
2 & 14 & 580.024 & 516.482 & 613.501 & 538.859 & 637.183 & 562.265 \\
2 & 15 & 612.022 & 540.797 & 647.314 & 572.300 & 675.967 & 604.029 \\
2 & 16 & 646.904 & 574.901 & 685.322 & 612.391 & 716.491 & 646.414 \\
2 & 17 & 981.976 & 609.741 & 722.939 & 649.241 & 755.304 & 682.219 \\
2 & 18 & 714.771 & 647.571 & 757.683 & 684.100 & 790.917 & 702.311 \\
2 & 19 & 749.324 & 679.309 & 793.296 & 703.049 & 827.350 & 735.388 \\
2 & 20 & 784.642 & 696.728 & 831.114 & 737.108 & 866.985 & 777.381 \\
2 & 21 & 818.136 & 728.585 & 868.816 & 774.163 & 907.360 & 807.014 \\
2 & 22 & 852.723 & 764.778 & 904.819 & 802.704 & 946.459 & 832.156 \\
2 & 23 & 888.137 & 792.796 & 941.137 & 828.950 & 984.102 & 871.453 \\
2 & 24 & 922.537 & 815.698 & 979.211 & 866.280 & 1021.99 & 914.490 \\
2 & 25 & 957.299 & 850.262 & 1017.29 & 905.561 & 1061.17 & 951.177 \\
2 & 26 & 992.689 & 887.230 & 1053.93 & 942.024 & 1101.43 & 976.345 \\
2 & 27 & 1027.87 & 923.354 & 1090.87 & 966.846 & 1141.57 & 1011.69 \\
2 & 28 & 1062.61 & 950.145 & 1129.15 & 998.759 & 1180.46 & 1053.86 \\
2 & 29 & 1098.14 & 975.476 & 1167.12 & 1038.43 & 1219.08 & 1088.51 \\
2 & 30 & 1133.72 & 1011.07 & 1204.56 & 1074.88 & 1258.10 & 1110.10 \\
\hline
           \end{tabular}
           }
   \caption{Period values for $\ell=1$ and $\ell=2$  modes corresponding to models with M$_*=$~0.338~M$_{\odot}$, effective temperature  of $10\, 000$~K and hydrogen envelope mass of 10$^{-4}$, 10$^{-5}$, and 10$^{-6}$~M$_{\odot}$.}     
    \label{apendix3}
\end{table}

\begin{table}
    \centering
        \resizebox{\columnwidth}{!}{%
    \begin{tabular}{ cc|cc|cc|cc }
    \hline
  \multicolumn{2}{c}{$M_H/M_{\odot}$} & \multicolumn{2}{c}{$10^{-4}$} & \multicolumn{2}{c}{$10^{-5}$} & \multicolumn{2}{c}{$10^{-6}$} \\
     \hline 
$\ell$ & k &  He &  C/O &  He & C/O & He & C/O \\
        \hline   
1 & 1  & 211.971 & 163.184 & 272.688 & 163.728 & 278.506 & 164.253 \\
1 & 2  & 276.632 & 198.670 & 286.577 & 253.161 & 349.212 & 303.724 \\
1 & 3  & 352.387 & 305.461 & 355.981 & 307.717 & 375.098 & 324.564 \\
1 & 4  & 418.508 & 365.398 & 439.196 & 385.267 & 440.661 & 386.324 \\
1 & 5  & 469.126 & 404.819 & 516.118 & 451.060 & 523.307 & 460.588 \\
1 & 6  & 526.313 & 461.883 & 573.907 & 497.554 & 604.686 & 542.008 \\
1 & 7  & 595.997 & 532.423 & 627.193 & 563.328 & 680.545 & 615.165 \\
1 & 8  & 657.552 & 589.357 & 695.003 & 637.210 & 738.264 & 665.314 \\
1 & 9  & 711.747 & 653.526 & 766.831 & 686.424 & 797.295 & 704.490 \\
1 & 10 & 780.216 & 697.500 & 833.735 & 731.373 & 868.130 & 768.783 \\
1 & 11 & 843.534 & 745.433 & 893.015 & 793.537 & 942.846 & 847.662 \\
1 & 12 & 901.897 & 808.247 & 961.966 & 869.583 & 1019.19 & 918.324 \\
1 & 13 & 968.255 & 875.514 & 1036.37 & 928.897 & 1094.93 & 970.636 \\
1 & 14 & 1035.46 & 925.732 & 1106.85 & 979.084 & 1164.70 & 1031.76 \\
1 & 15 & 1097.32 & 977.308 & 1172.29 & 1045.33 & 1233.01 & 1100.78 \\
1 & 16 & 1162.30 & 1043.90 & 1241.03 & 1107.19 & 1306.57 & 1139.70 \\
1 & 17 & 1230.74 & 1099.27 & 1314.57 & 1145.56 & 1383.92 & 1213.45 \\
1 & 18 & 1296.20 & 1132.24 & 1387.83 & 1214.51 & 1462.30 & 1297.58 \\
1 & 19 & 1360.81 & 1197.61 & 1458.53 & 1291.93 & 1539.71 & 1374.64 \\
1 & 20 & 1429.04 & 1270.06 & 1527.91 & 1369.63 & 1614.85 & 1424.85 \\
\hline
2 &  1 & 122.673 & 111.022 & 161.731 & 111.709 & 177.473 & 112.048 \\
2 &  2 & 177.534 & 115.239 & 179.106 & 146.305 & 211.221 & 178.889 \\
2 &  3 & 219.601 & 181.074 & 224.211 & 182.664 & 226.761 & 188.896 \\
2 &  4 & 254.109 & 214.244 & 273.063 & 228.523 & 274.394 & 229.265 \\
2 &  5 & 283.806 & 237.817 & 313.728 & 263.906 & 322.327 & 270.651 \\
2 &  6 & 320.563 & 270.466 & 343.764 & 289.721 & 369.126 & 316.276 \\
2 &  7 & 361.292 & 310.114 & 378.605 & 328.255 & 408.989 & 358.240 \\
2 &  8 & 392.415 & 342.644 & 419.762 & 372.913 & 440.332 & 391.682 \\
2 &  9 & 428.280 & 384.076 & 460.811 & 410.346 & 478.305 & 423.698 \\
2 & 10 & 467.738 & 417.823 & 496.879 & 433.479 & 507.736 & 451.141 \\
2 & 11 & 502.082 & 439.454 & 532.901 & 463.571 & 564.517 & 493.970 \\
2 & 12 & 537.746 & 471.766 & 575.233 & 507.100 & 608.865 & 537.519 \\
2 & 13 & 577.290 & 510.640 & 617.606 & 544.610 & 651.321 & 570.193 \\
2 & 14 & 614.507 & 543.309 & 656.751 & 574.361 & 690.583 & 602.753 \\
2 & 15 & 650.556 & 573.042 & 694.888 & 610.217 & 731.419 & 647.504 \\
2 & 16 & 689.501 & 608.268 & 736.295 & 655.098 & 775.176 & 697.196 \\
2 & 17 & 728.692 & 646.436 & 779.002 & 698.230 & 820.347 & 733.340 \\
2 & 18 & 765.618 & 688.715 & 820.721 & 730.570 & 865.547 & 756.089 \\
2 & 19 & 804.125 & 724.033 & 861.022 & 752.829 & 909.721 & 799.923 \\
2 & 20 & 843.683 & 741.825 & 901.474 & 797.069 & 952.601 & 844.521 \\
2 & 21 & 882.106 & 778.876 & 943.447 & 839.523 & 995.140 & 871.128 \\
2 & 22 & 920.408 & 820.648 & 986.213 & 865.190 & 1038.45 & 909.888 \\
2 & 23 & 959.520 & 851.564 & 1028.20 & 898.880 & 1082.74 & 959.228 \\
2 & 24 & 998.867 & 876.304 & 1069.13 & 645.024 & 1127.70 & 1003.20 \\
2 & 25 & 1037.62 & 916.100 & 1110.60 & 988.842 & 1172.86 & 1032.01 \\
2 & 26 & 1076.42 & 958.321 & 1153.10 & 1019.67 & 1217.46 & 1070.84 \\
2 & 27 & 1116.01 & 995.860 & 1195.53 & 1052.01 & 1261.71 & 1117.93 \\
2 & 28 & 1155.22 & 1022.68 & 1237.56 & 1095.90 & 1305.52 & 1157.13 \\
2 & 29 & 1194.12 & 1057.21 & 1279.30 & 1137.63 & 1349.21 & 1187.67 \\
2 & 30 & 1233.58 & 1097.80 & 1321.30 & 1166.17 & 1393.53 & 1231.34 \\
\hline
\end{tabular}
}
   \caption{Period values for $\ell=1$ and $\ell=2$  modes corresponding to models with M$_*=$~0.338~M$_{\odot}$, effective temperature of \textbf{ $9\, 000$}~K and hydrogen envelope mass of 10$^{-4}$, 10$^{-5}$, and 10$^{-6}$~M$_{\odot}$.}     
    \label{apendix4}
\end{table}


\bsp	
\label{lastpage}
\end{document}